\documentclass[11pt]{article}
\pdfoutput=1
\usepackage{graphicx}
\usepackage{hyperref}
\usepackage{amssymb}
\usepackage{amsmath}

\newcommand{\lam}		{\lambda}
\newcommand{\vep}		{\varepsilon}
\DeclareMathOperator{\sign}{sign}

\def \om{\omega}

\def \ha{\frac{1}{2}}		
\title{Interacting Electrons in One Dimension: The Two Chain Perylene-Metal Dithiolate Series}
\author{L.~Alc\'{a}cer$^1$,\ \ R.T.~Henriques$^1$ and M.~Almeida$^2$\\
{$^1$ \small Instituto de Telecomunica\c{c}\~{o}es, Lisboa, Portugal}\\ 
{$^2$ \small C2TN, P\'{o}lo de Loures do IST, Bobadela, Portugal}\\[1mm]\\
\scriptsize{To appear in ''Essays in Honour of Am\'{i}lcar Sernadas'', College Publications, London, 2017}}

\date{} 
\begin{document}
\maketitle
\section*{Overview}
In 1964, William Little, from Stanford University,  proposed a model to synthesize a room temperature organic superconductor \cite{Little:1964aa}, which, a few years later,  led to a new field of research entitled \emph{one-dimensional conductors}. In those materials, the electrical conductivity is high along one direction and much lower in the other directions. This is possible in two cases: i) in solids in which flat organic molecules are stacked on top of each other, and through intermolecular interactions, electrons can travel along the staking axis, and ii) in organic polymers in which single carbon-carbon bonds alternate with double bonds allowing electrons to travel along the polymer chain. 
Nowadays, one can confine electrons in one or two dimensions by means of  top-down approaches to nanotechnology, but that is not the subject of the present paper. The present paper deals with organic crystalline solids, with extreme anisotropy, in which the one-dimensional character of the interactions of the electrons with each other and with the lattice vibrations (the phonons) are the source of various types of competing instabilities and phase transitions. Take, for example, a linear chain of $N$ sites, separated from each other by a distance $a$, the \emph{unit cell} size, and with one electron per site, or unit cell.  Since each site can accommodate two electrons, with opposite spins, the system would have $2N$ states available, but only $N$ would be occupied. One would have, therefore, a \emph{half-filled band} metal. However, the total energy of the electrons would decrease if the sites would arrange themselves in pairs, doubling the unit cell, which correspond to opening a gap at the Fermi level, the last occupied energy level.   
In one-dimension, below a certain critical temperature, the energy needed to distort the lattice, doubling the unit cell, becomes lower than the \emph{gain} in energy of the electrons, and a phase transition may occur. This is the Peierls transition, one of many that may occur. 

A good example of a one-dimensional solid is  the one, best known as (TMTSF)$_2$PF$_6$, which was the first organic superconductor to be discovered, in 1979, and has been a centre of attention since then. In the words of Paul Chaikin, a well known physicist, now at the New York University, this is  ''the most remarkable electronic material ever discovered. In one single crystal, one can observe all the usual competitions, all electron transport mechanisms known to man, plus somethings new''.

In an instance of lucky serendipity, one of the present authors (LA), synhtesized the first members of a family of organic conductors based on the perylene molecule and metal-dithiolate complexes \cite{Alcacer:1970aa} (see Figure \ref{Fig_1}). At the time,  it was not possible to prove that any of these materials was a metal and much less a superconductor, in spite of the many evenings spent cooling them down to liquid helium temperature. They always turned insulators! The first papers with a full description of their electrical and magnetic properties  appeared only in 1974 and 1976 \cite{Alcacer:1974aa,Alcacer:1976aa} as the outcome of more detailed studies. 
This family of materials of general formula (Per)$_2$[M(mnt)$_2$], with M = Pt, Au, etc., became of considerable interest later on for exhibiting metallic behaviour and very high one-dimensionality. The crystal structures are formed by chains of  [M(mnt)$_2$] units surrounded by perylene chains, which are 3/4 filled band metals and therefore highly conducting at room temperature. On the other hand, the [M(mnt)$_2$] chains are insulating, with localized spins  in the case of M=Ni and Pt, (S=1/2), for example, spins that \emph{exchange fast}, with the itenerant electrons on the perylene chains. For more than 40 years, now, these materials have also been amazing us, with their remarkable properties, some of them typical of the extreme anysotropy, but mostly not expected as they came. The nature of the metal-insulator transition in the Pt compound, for example, is particularly compelling. At the temperature of 8~kelvin, the perylene chain tetramerizes, developing a \emph{charge density wave} (CDW), with the opening of a gap. This lattice distortion drives a distortion of the [M(mnt)$_2$] chains, which dimerize, leading to a spin-Peierls ground state (SP). The spin-Peierls state in the [M(mnt)$_2$] chains is coupled to the charge density wave state in the perylene chain, up to magnetic fields of the order of 20 tesla, a situation not foreseen in the present theory, which predicts a decoupling at much lower fields. Under extreme conditions of low temperature, high pressure and high magnetic fields, these materials exhibit a rich variety of phenomena, including field induced charge density waves (FICDW), interference effects of electrons from "trajectories" on different Fermi surface sheets, and for our delight, superconductivy was finally observed, under pressure, in 2009, evolving in a rather unusual way, from the charge density wave state. This paper is about the wonderful materials of this family.
%
\section{Introduction} 
On page 111, of his 1955  ''Quantum Theory of Solids'', \cite{Peierls:1955aa}, Rudolf Peierls states that ''it is ... likely that a one-dimensional model could never have metallic properties''. This he concludes, after arguing that in a 1D metal with a partly filled band the regular chain  could not be stable, since one would always find a distortion corresponding to a certain number of atoms in the unit cell for which a break would occur at or near the edge of the Fermi level \cite{Frohlich:1954aa}. 

In the early 1960's, a few years after the success of the BCS theory of superconductivity \cite{Bardeen:1957aa}, there was the hope to learn how to produce materials which could be superconductors at room temperature. In 1964, William Little's seminal paper entitled ''Possibility of Synthesizing an Organic Superconductor''~\cite{Little:1964aa} stimulated the search for new materials which could comply with his model system. Based on London's idea \cite{London:1950aa}  that superconductivity might occur in organic macromolecules, he proposed a mechanism similar to the BCS theory of superconductivity and even suggested model molecules that could lead to room temperature superconductivity. His model structure was composed of a conjugated polymer chain dressed with highly polarizable molecules as side groups. The polymer chain would be a one-dimensional metal with a single mobile electron per C-H unit; electrons on separate units would be Cooper-\emph{paired} by interacting with the exciton field on the polarizable side groups. However, in spite of much effort, nobody succeeded in making Little's molecule, nor room temperature organic superconductors! 

The first organic conductor, a perylene-bromine complex, had been reported in 1954 by H. Akamatu, H. Inokuchi and Y. Matsunaga \cite{Akamatu:1954aa}. As many important discoveries, this was a case for serendipity. At the time, nobody would believe that an organic substance could ever be a conductor. Linus Pauling, who was visiting Japan at the time, was truly surprised.  In the late 1960's and early 1970's, several organic semiconductors were reported and the subject became of much interest, especially due to William Little's paper.

\begin{figure}[htb]
\centerline{\includegraphics[width=0.65\textwidth]{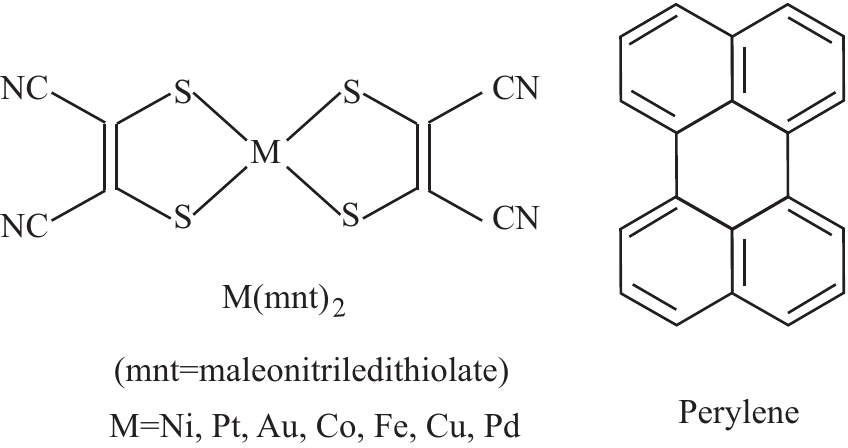}}
\caption{Constituent molecules of the perylene-metal dithiolate charge transfer salts.} 
\label{Fig_1}
\end{figure}

Before going into more detail on the (Per)$_2$[M(mnt)$_2$] series, (see Figure~\ref{Fig_1}), it should  be mentioned that, in the meanwhile, several organic conductors were sinthesized, in particular TTF-TCNQ, a charge-transfer salt of tetrathiafulvalene (TTF) and tetracyanoquinodimethane (TCNQ)  reported to exhibit \emph{superconducting fluctuations and Peierls instability} \cite{Coleman:1973aa}. Superconductivity was not confirmed in this compound, but the discussion it triggered led to considerable experimental and theoretical developments, and the search for organic superconductivy was on its way. 

At last, in 1979, the first organic superconductor was synthesized by Klaus Bechgaard  and found to have a transition temperature of $T_c = 1.1$~K, under an external pressure of $6.5$ kbar \cite{Jerome:1980aa}. Such finding gave rise to the dawn of a new era in the field of organic  conductors and superconductors. Several series of organic superconductors were synthesised and the needed theory  was developed.

Little's idea was based on conjugated polymers, for which single carbon-carbon bonds alternate with double bonds, and, in principle, they should be metals or semiconductors, depending on the unit cell. Polyacetylene was the paradigm compound and had been under investigation for some time  \cite{Shirakawa:1971aa}, but it was shown to become highly conducting only through heavy doping \cite{Shirakawa:1977aa}. That accidental discovery led to the thorough study of conducting polymers, which were the subject of the 2000 Nobel Prise in Chemistry awarded to Alan Heeger, Alan MacDiarmid and Hideki Shirakawa. A new technology emerged and is now the important field of \emph{Organic Electronics}. 

\section{The Interesting Physics: 1 D Instabilities}
In a one-dimensional crystal, fluctuations destroy all long-range order, and since there are always fluctuations at any finite temperature, such a system cannot be ordered except at zero temperature. 
This makes the 1D real systems interesting---although not strictly 1D, they exhibit many interesting kinds of fluctuations, instabilities and phase transitions.   
This is what happens in solids with extreme anisotropy, as in most organic conductors and superconductors, where the electron-electron interactions mediated by phonons originate structural instabilities and phase transitions, namely of the metal-insulator and superconducting type. The needed theoretical framework is identical to that of superconductivity and all these \emph{critical phenomena} benefit of a common formalism. 
The anisotropy in the properties of these solids, such as the electrical conductivity along the chain, which can attain values as high as  $10^5$ times the value in the perpendicular directions,  is due to the highly directional $\pi-\pi$ interactions along the stacks of flat molecules or along the chains of carbon atoms in conjugated polymers.

\begin{figure}[htbp]
\centering
 \includegraphics[scale=0.9]{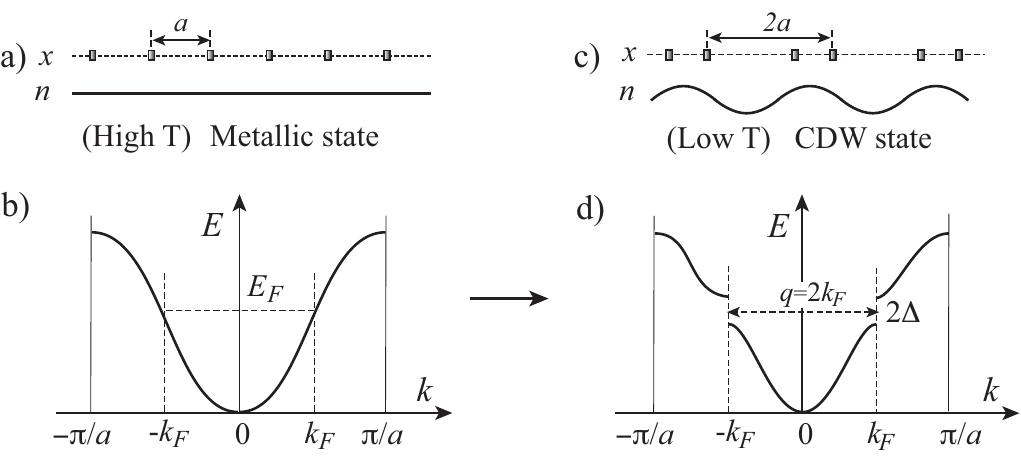}  
 \caption{a) 1 D lattice, electron density {$n$},  corresponding to one electron per site, and b) electron dispersion on the regular chain with a half filled band. c) and d) Peierls modulated in the electron-phonon system. States with {$|k|\lesssim k_F$} gain energy by the {$q=2k_F$} distortion (lattice doubling), leading to the opening of an energy gap at {$k_F$}.}
\label{Fig_2}
\end{figure}

One can consider several types of instabilities in a 1 D crystal, namely:
\begin{enumerate}
\item In the presence of electron-phonon interactions, the ground state is unstable relative to the development of \emph{charge density waves} (CDW) with wave vector $q$ twice the Fermi wave vector ($q=2k_F$). This is the \emph{Peierls instability} and  competes, in general,  with the BCS superconducting instability.
\item In the presence of magnetic interactions, the ground state is unstable relative to the development of \emph{spin density waves}, (SDW), without lattice modulation, and \emph{spin-Peierls instabilities}, when there is a lattice distortion. 
\item In a 1 D system with short range interactions, the thermal fluctuations destroy the long range order at any temperature $T>0$.
\item In a 1 D system,  an arbitrarily small disorder induces electron localization and the consequent transition from a metal into an insulator. 
\end{enumerate}
Figure \ref{Fig_2} illustrates how, due to the electron-phonon interaction, a regular linear chain of molecules with one electron per molecule, in a metallic state, (half filled band system) undergoes a Peierls distortion to a \emph{charge density wave} (CDW) state, by doubling the size of the unit cell, and becomes an insulator. 

In Figure \ref{Fig_2}~a), a chain of molecules along the $x$-axis, equally spaced by a lattice parameter $a$, with one electron per unit cell, is a metal at high temperature with constant electron density, $n$.   If there is one electron per unit cell in the high temperature state, the energy band is half-filled and $k_F=\pi/2a$, as shown in Figure \ref{Fig_2}~b).  At low temperature, the lattice parameter is modulated and tends to double due to the electron-phonon interaction, and a \emph{charge density wave} (CDW) is generated (Figure \ref{Fig_2}~c)). Whenever the number of electrons per unit cell is less than two, the band is partially filled and an identical situation can occur. In general, if the degree of band filling is $1/n'$, $k_F=\frac{1}{n'}\frac{\pi}{a}$ and the stable charge density wave will have a wave length $n'a$. However, we should take into account that, in a band more than half filled, \emph{holes} should be considered instead of electrons, and the degree of filling would be $1-n'$. A $3/4$ filled band would correspond to a $1/4$ \emph{hole} filled band, and the charge density wave (of positive charge) would have a wave length $4a$, giving rise to  a tetramerization---see below. Since in one dimension the elastic energy needed to modulate the crystal lattice is less than the energy gain in the conduction electrons, when the temperature is lowered, the opening of the gap, $2\Delta$, at $k_F$, drives a structural instability and eventually a structural phase transition. 
%
\subsection{The Peierls Instability}
As mentioned before, Peierls \cite{Peierls:1955aa} and Frohlich \cite{Frohlich:1954aa} predicted that a 1 D electron system in a deformable lattice is unstable relative to a modulation of the lattice with wave vector $q=2k_F$, in which  $k_F$ is the Fermi wave vector. For an overview of the theoretical background we will follow the excellent theoretical introduction of D. J\'{e}r\^{o}me and H. J. Schulz \cite{Jerome:2002aa} to the  instabilities of quasi-one-dimensional electron systems originally published in 1982 in \textit{Advances in Physics} and which should be referred to for more detail.    
A one-dimensional electron  system can be described by the following Hamiltonian, in second quantization, 
\begin{equation}\label{eq:Jerome1}
H=\sum_{ks} \vep_k\ a^+_{ks}a_{ks}+\sum_q\om_q\ b^+_qb_q+\frac{g}{\sqrt{L}}\sum_{kqs}a^+_{k+q,s}a_{ks}\left(b_q+b^+_{-q}\right) \end{equation}
where $a^+_{ks}$ and $a_{ks}$ are the creation and annihilation operators for an electron in state $k$, with spin $s$, and energy $\vep_k$; $b^+_{q}$ and $b_{q}$ are the creation and annihilation operators for a phonon with wave vector $q$ and energy $\om_q$; $L$ is the length of the system. For simplicity we make $\vep_{k_F}=0$ and $\hbar=1$.
The first two terms describe the non-interacting electron and phonon systems, respectively, and the third term is the electron-phonon interaction, with coupling constant $g$, usually a function of $k$ and $q$. In the present context, electrons with $|k|\approx k_F$ and phonons with $|q|\approx2k_F$ are the most important and, therefore, the wave vector dependence on  $g$ can be ignored. The sums are limited to the Brillouin zone.

The lattice oscillations $u$ can be written in terms of the phonon operators, as
\begin{equation}\label{eq:Jerome2}
u(x)=\sum_q\frac{1}{\sqrt{2L\om_q}}  \left(b_q+b^+_{-q}\right) e^{iqx}
\end{equation}
A modulation of the lattice with wave vector $q=2k_F$ is described by \(\langle b_{2k_F}\rangle=\langle b^+_{-2k_F}\rangle \propto \sqrt{L} \), \textit{i.e.}, the phonon modes with wave vector $\pm k_F$ are occupied.

In the mean field approximation, one can approximately describe the system by the Hamiltonian 
\begin{equation}\label{eq:Jerome3}
\begin{split}
H=& \sum_{ks}\vep_ka^+_{ks}\,a_{ks}+ 2\om_{2k_F}\left|\langle b_{2k_F}\rangle\right|^2 +  \\
  &+\frac{2g}{\sqrt{L}}\sum_{ks}\left(a^+_{k+2k_F,s}a_{ks}  \langle b_{2k_F}\rangle +a^+_{k-2k_F,s} a_{ks}\langle b^+_{2k_F}\rangle \right) 
\end{split}
\end{equation}
this meaning that the lattice modulation mixes the $k$ states with the $k\pm 2k_F$ states. More important is the mixing of the quasi-degenerate states with $|k|\approx k_F$ that can be approximately treated in the free electron model and by linearizing $\vep_k$ in the vicinity of the Fermi points, \textit{i.e.}, \( \vep_k=v_F(|k|-k_F)\),  $v_F$ being  the Fermi velocity, giving
\begin{equation}\label{eq:Jerome4a}
H=\sum_{ks} E_k\, c^+_{ks}c_{ks}+  L\frac{\om_{2k_F}}{2g^2}|\Delta|^2  
\end{equation}
with
\begin{equation}\label{eq:Jerome4b}
E_k= \sign \left(|k|-k_F\right) \left[v^2_F\left(|k|-k_F\right)^2+|\Delta|^2\right]^{1/2}    
\end{equation}
$c^+_{ks}$ and $c_{ks}$ are now the creation and annihilation operators for electrons in states which are linear combinations of the old states $k$ e $k\pm2k_F$, and in which we define the \emph{order parameter}
\begin{equation}\label{eq:Jerome5}
\Delta=\frac{2\,g\,\langle b_{2k_F}\rangle}{\sqrt{L}}
\end{equation}
As it can be seen  in Figure \ref{Fig_2}~d), the energy of the electrons is lowered and  the lattice modulation entails the opening of a gap of width $2|\Delta|$ at the Fermi level, as shown from equation  (\ref{eq:Jerome4b}).  At the same time, the elastic energy increases due to the modulation of the lattice described by the second term of equation (\ref{eq:Jerome4a}).

The overall effect on the total energy of the electronic system can be calculated to give  
\begin{equation}\label{eq:Jerome6}
\begin{aligned}
 E_{el}(\Delta)	&=2\sum_{k=-k_F}^{k_F} E_k   \\
			&=  -\frac{L\,n\,E_F}{2}\left[ 1+\frac{|\Delta|^2}{E_F^2}\ln\left(\frac{2\,E_F}{|\Delta |}\right) + \text{higher order terms}\right] 
\end{aligned}
\end{equation}
 In this equation,  $E_F$  is the Fermi energy, measured relative to the bottom of the band, and $n$ is the electron density. From equation (\ref{eq:Jerome6}), one can see that the energy gained by the electrons is, for small $|\Delta|$, proportional to  $-|\Delta|^2\ln |\Delta| $, which is always bigger than the loss in elastic energy which is only proportional to $|\Delta|^2$. From this, one can conclude that, in the mean field approximation, \emph{the system is unstable relative to a modulation of the lattice, for an arbitrarily small coupling}.
 To get the value for the gap, one can minimize the energy and obtain 
\begin{equation}\label{eq:Jerome7}
|\Delta|=2\,E_F\,e^{-1/\lam}, \qquad \lam=\frac{2\,g^2}{\pi\,v_F\,\om_{2k_F}}
\end{equation}
where $\lam$ is the wave length of the lattice modulation.
It should be noted that the logarithmic term in  (\ref{eq:Jerome6}) only appears if the gap opens exactly at 
$\pm k_F$. In other words, the wave length of the lattice modulation is determined by the electronic band filling and is equal to $\pi/k_F$. The expectation value of the modulation in the ground state  is 
\begin{equation}\label{eq:Jerome8}
\langle u(x)\rangle=\sqrt{\left(\frac{2}{\om_{2k_F}} \right)}\frac{|\Delta| }{g}\cos(2k_F\,x+\phi)   
\end{equation}
where   $\phi$  is the phase of  $\Delta$ (\textit{i.e.}, $| \Delta | e^{i\phi}$): there is a modulation of the electron density as a consequence of the lattice modulation, usually known as a \emph{charge density wave}, (CDW).

The excited states depend only on the amplitude of $\Delta$ and not on the phase. Due to the translational symmetry, the CDW can move along the crystal, transporting a d.c. current, which ideally would be infinite, but which is suppressed in real crystals due to several mechanisms such as impurities and lattice locking.

The description of the Peierls ordered state for $T=0$ is qualitatively valid for temperatures near zero, while the long range order is not destroyed. At sufficiently high temperatures, the complete destruction of the order is expected, and the thermally excited electrons  above the gap have energies $\approx |\Delta|$ and will gain energy if the gap decreases. A decrease of  $|\Delta|$ enables the additional excitation through the gap, and the mechanism predicts  the disappearance of the gap and of the long range order above a certain temperature.
 
\begin{figure}[htbp]
\centering
 \includegraphics[scale=0.9]{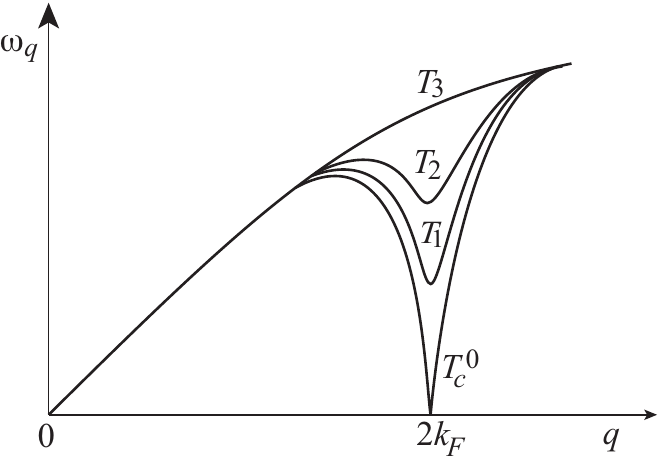}
 \caption{Evolution of the Kohn anomaly, in the phonon spectrum, with the temperature {($T_3>T_2>T_1>T^0_c$)}.}
 \label{Fig_3}
\end{figure}

In these 1 D systems, a sharp dip in the phonon spectrum, at the wave vector $q=2k_F$, known as the Kohn anomaly \cite{Kohn:1959aa},  is particularly noticeable (Figure \ref{Fig_3}). It is due to the possibility of exciting all the electrons from one side of the Fermi distribution into the other side with the single wave vector and very little energy. 

The phonons with wave vector near $q=2k_F$ involve displacements of the lattice and are collective oscillations strongly coupled to the lattice, as well as to the charge density. It is this coupling that is responsible for the dip in the phonon spectrum. As the temperature is lowered, the dip increases and, at the critical temperature   $T_c^0$, one finally has $\om_{2k_F}=0$, indicating the instability of the system relative to a modulation with wave vector $q=2k_F$---the complete softening of the lattice and a static distortion appears. This is one explanation for charge density waves in solids. The wave vectors at which a Kohn anomaly is possible are the nesting vectors of the Fermi surface, that is, vectors that connect most of the points of the Fermi surface. For a one-dimensional chain of atoms, this vector would be $2 k_F$, connecting entirely the two flat Fermi Surfaces.
 
The description of the Peierls transition is very similar  to the BCS theory of superconductivity, namely:
\begin{itemize}
\item In both cases, the gap opens on the whole Fermi surface inducing a rapid decrease in the energy of the electrons.       
\item In both cases, the instability is expressed by a correlation function which diverges logarithmically.        
\item The transition temperature and the gap at $T=0$ only differ by a proportionality factor, following the equation {\(2|\Delta_{(T=0)}|/T_c=3.5\)}.
\item The effect of impurities in the  Peierls transition is analogous to the effect of magnetic impurities in a superconductor. 
\end{itemize}
%
\subsection{Competing Instabilities in a 1 D Electron Gas}
In the framework of mean-field theory, the electron-electron interaction near the Fermi level can be characterized by four coupling constants, $g_1, g_2, g_3$ and $g_4$, their values determining the most stable phases.
$g_1$ corresponds to the electron-electron interaction with momentum transfer $\pm 2k_F$ (backward scattering),  $g_2$  and $g_4$ have small momentum transfer (forward scattering) involving electrons at both $k_F$ and $-k_F$, ($g_2$), or only on one side, ($g_4$). For $g_3$, the total electron momentum is changed by $4k_F$, being only allowed if $4k_F$ is a reciprocal lattice vector (umklapp scattering), \textit{i.e.} for a half-filled band.

Just to have a qualitative notion of the problem, one can, in the mean-field approximation, look for the possibility of opening a gap at the Fermi level, by evaluating the expectation values of some operators which may originate stable ordered states. 
Let us then consider the following operators  
 \begin{align}
O_{CDW}(q)&	=\sum_{ks}b^+_{-k_F+k-q,s} \, a_{k_F+k,s}    	\label{eq:Jerome9}\\
	O_{SDW\alpha}(q)&=\sum_{kss'}b^+_{-k_F+k-q,s}\, \sigma_{\alpha}^{ss'} \,a_{k_F+k,s'}  \quad (\alpha=x,y,z)  \label{eq:Jerome7a}\\
	O_{SS}(q)&=\sum_{ks}  b^+_{-k_F-k,-s}\,  a_{k_F+k+q,s}  	\label{eq:Jerome8a}\\
	O_{TS\alpha}(q)&=\sum_{kss'} b^+_{-k_F-k,-s}\, \sigma_{\alpha}^{ss'}  \,a_{k_F+k+q,s'}   \quad (\alpha=x,y,z)	\label{eq:Jerome9c} 
\end{align}
where $a$ and $b$ are operators which describe electrons moving to the right and to the left, respectively, and 
$\sigma_{\alpha}^{ss'}$ are elements of the spin Pauli matrices. The operators $O_{CDW}(q)$ and $O_{SDW_\alpha}(q)$ are the Fourier components of the charge and spin densities, with wave vector  $2k_F+q$, respectively. The  $O_{SS}(q)$ and  the three possible $O_{TS_\alpha}(q)$ are the \emph{singlet} and \emph{triplet} operators of the Cooper pairs of the superconducting state.   To look for ordered states, one would have to compute the expectation values of these operators. If, for example, one considers the \emph{singlet} superconducting state, the Hamiltonian leads to a solution which is similar to the Peierls transition with a gap $|\Delta_S|$ in the Fermi surface which opens at $q=0$.
\begin{equation}
\begin{aligned}
	|\Delta_{SS}|&=2\,E_F \,e^{-\frac{1}{\lam_{SS}}}\\
	\lam_{SS}&=\frac{g_1+g_2}{2\pi\,v_F}
\end{aligned}
\end{equation}
for  $g_1 + g_2 < 0$. For  $g_1 + g_2 > 0$, there is no self-consistent solution with finite $|\Delta|$.

\begin{figure}[htbp]
\centering
 \includegraphics[scale=0.9]{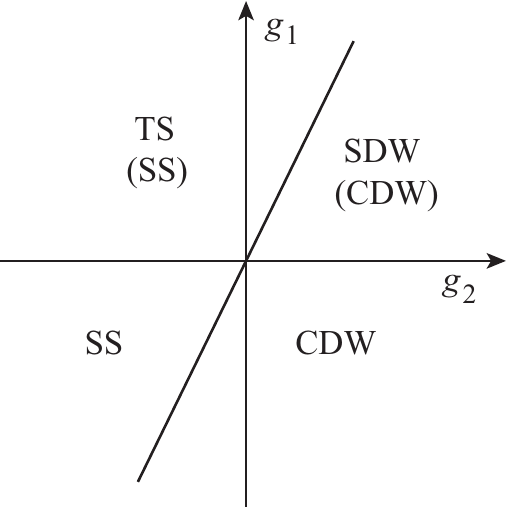}
 \caption{Phase diagram in the {$g_1-g_2$} plane showing the domains for which the various operators give stable phases.}
 \label{Fig_4}
\end{figure}

The transition temperature is given by
\begin{equation}
T_c=\frac{2C}{\pi}E_F\,e^{\frac{1}{\lam_{SS}}} \ \ ,  \qquad (C=1.781)
\end{equation}
The same argument can be applied to the other phase transitions,  replacing  $\lam_{SS}$ by the respective $\lam$ parameters:
\begin{align}
\lam_{CDW} & =-\frac{2g_1-g_2}{2\pi\,v_F}\\
\lam_{SDW\alpha} & =\frac{g_2}{2\pi\,v_F}\\
\lam_{TS\alpha} & =-\frac{g_1-g_2}{2\pi\,v_F}
\end{align}
Figure  \ref{Fig_4} schematically represents the domains in the $g_1-g_2$ plane for which the various $O$ operators give stable phases.

The more stable phase is the one which will have the highest transition temperature. Besides the CDW of Peierls type all other phases are possible depending on the $g_1, g_2, g_3, g_4$ coupling constants. The superconducting \emph{singlet}   (SS, CDW) and \emph{triplet} (TS, SDW) occur for $g_1<0$ and $g_1>0$ respectively. The density wave and superconducting phases are separated by the line $g_1=2g_2$ (Fig.\ref{Fig_4}).
%
\section{The  Perylene-Metal Dithiolate Series}  
\subsection{Crystal Structure, Energy Bands and Fermi Surface}
This series of charge-transfer salts appears in several phases with different crystal structures. There are also perylene-metal dithiolate salts with other stoichiometries. In this paper, we will focus on the so called $\alpha$-phases which are the most studied and, so far, the most interesting. 

\begin{figure}[htbp]
\centering
\includegraphics[width=7.5cm]{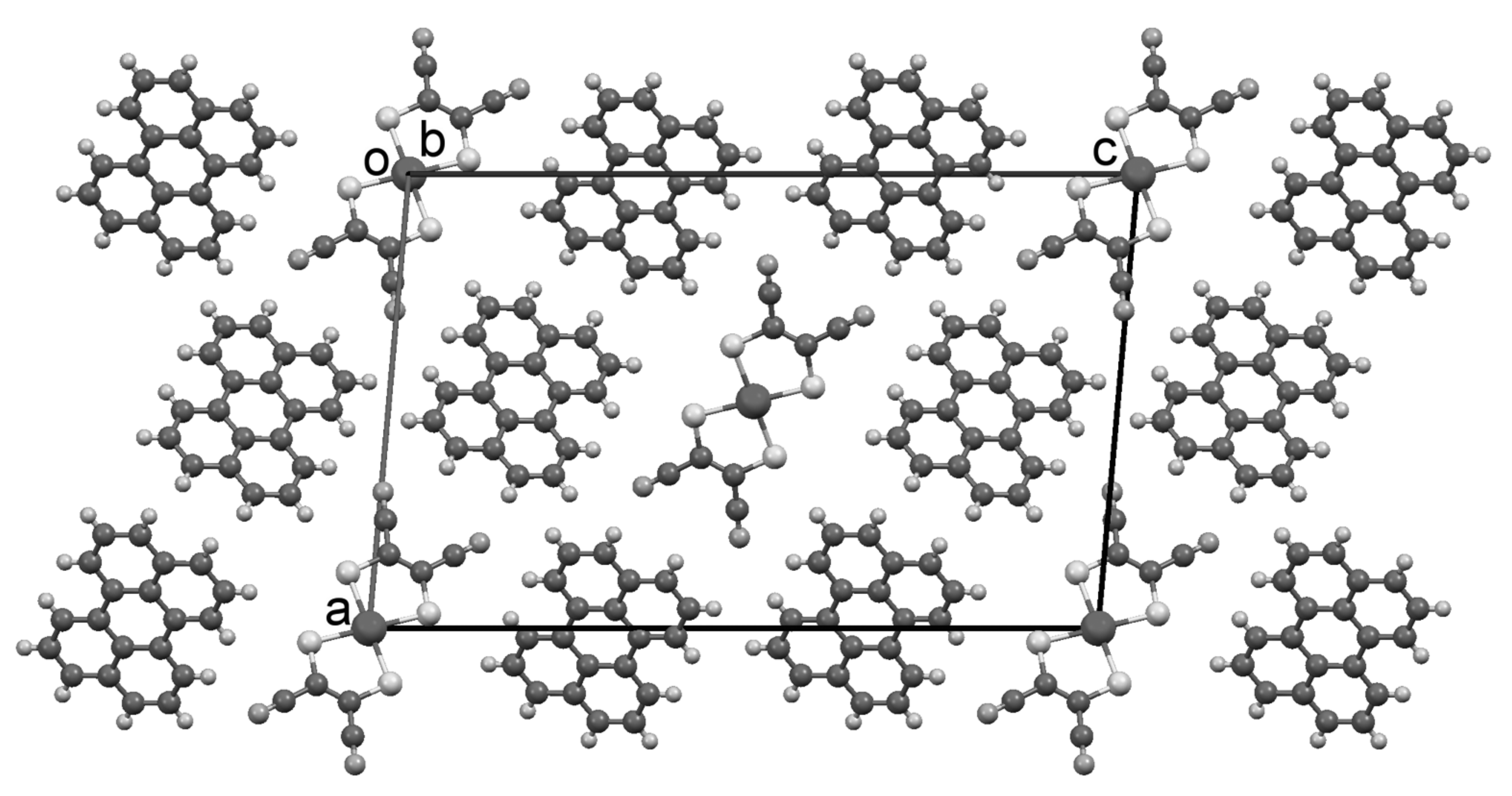}  
\caption{Crystal Structure of the {$\alpha$}-phase of the {(Per)$_2$[Pt(mnt)$_2$]}.  Projection along the {$b$} axis.}
\label{Fig_5}
\end{figure}

The $\alpha$-(Per)$_2$[M(mnt)$_2$], with M=Ni, Au, Cu Pt or Pd, form an isostructural series, crystallizing in the monoclinic system (space group  P2$_1$/c)  \cite{Alcacer:1980aa,Henriques:1984aa,Domingos:1988aa}. The Fe and Co compounds are dimerized, the $b$ axis being doubled. The projection of the crystal structure of the platinum compound along the $b$ axis is illustrated in Fig.\ref{Fig_5}. 
%
It shows that there are segregated regular stacks of [Pt(mnt)$_2$] and perylene parallel to $b$. Each column of [Pt(mnt)$_2$] is surrounded by six columns of perylene and each perylene stack has three stacks of perylene and three stacks of anions as nearest neighbours. The dihedral angle between the two mean planes of the molecules is $67^\circ$. Interchain coupling is weak, with  intermolecular distances  only  slightly smaller than the sum of the van der Waals radii of the nitrogen and carbon atoms of the two molecules. In the stack, adjacent overlapping  [Pt(mnt)$_2$]  anions  are separated by $3.65$~\AA,\  the distance between consecutive Pt atoms is $4.19$~\AA \ and the tilt of the plane with respect to $b$ is $29^\circ$. The perylene molecules present a very good graphite-like mode of overlap at the distance of $3.32$~\AA, shorter than in graphite ($3.35$~\AA), the tilt of the plane with respect to $b$ being $37^\circ$. This mode of stacking allows very short distances between carbon atoms ($\approx 3.3$~\AA) \ and therefore there is consistent evidence for strong interactions.
In table \ref{tabela1}, the cell parameters for the different (Per)$_2$[M(mnt)$_2$] compounds are compaired \cite{Domingos:1988aa,Gama:1991ab}. 

 \begin{table}[htp]
\caption{Cell parameters for {(Per)$_2$[M(mnt)$_2$]} \cite{Domingos:1988aa,Gama:1991ab}. In the Co and Fe salts, the $b$ axis is doubled, when compared with the others \cite{Gama:1991ab}.}
\begin{center}
\begin{tabular}{c|ccccccc}
\hline 
 M & Au & Pt &  Pd & Ni  & Fe & Co\\
\hline 
$a$~(\AA) 	& 16.602 & 16.612 & 16.469 & 17.44 &  17.600 & 17.75 \\  
$b$~(\AA) 	& 4.191	& 4.194 	& 4.189 	& 4.176 & 8.136 & 8.22 \\
$c$~(\AA)		& 30.164 	& 30.211 & 30.057 	& 25.18 & 30.088 & 30.88\\
$\beta~(^\circ)$  	& 118.69	&118.70 & 118.01 	& 91.57 & 123.72 & 123.0\\
\hline   
\end{tabular}
\end{center}
\label{tabela1}
\end{table}
For the $\alpha$-phases of  (Per)$_2$[M(mnt)$_2$], the energy band in the perylene chain is quarter filled and the band in the [M(mnt)$_2$] chain is half filled. The perylene chain is metallic at room temperature and undergoes a metal-insulator transition at low temperature. The [M(mnt)$_2$] chain, for M=Pt, Ni, Pd (S=1/2), behaves as a Mott-insulator at room temperature \cite{Alcacer:1980aa} and undergoes a spin-Peierls transition, at low temperature. 
 
The band structure of the perylene chain was first calculated, by Veiros \textit{et al}  \cite{Veiros:1994aa} and more recently by Canadell \textit{et al} \cite{Canadell:2004aa} (using the extended Huckel approach with both single-$\zeta$  and double-$\zeta$ type atomic basis sets). These calculations estimate the intrachain transfer integral ($t_b$) as 148.6~meV (or 353.8~meV for the double-$\zeta$ basis set), with the other transfer integrals in the range of 0 - 2.4~meV (or  0 - 8~meV for the double-$\zeta$). This confirms that these materials are nearly perfect one-dimensional conductors, and that the conduction is mainly along the stacking direction of the perylene molecules, which are partially oxidized,~ (Per)$^{+1/2}$.   These values of the transfer integrals  are in agreement with the experimental values of the bandwidth, $W$, which are in the range 0.4 - 0.6~eV ($W=4t$) depending on the metal M, as estimated from both thermopower and magnetic susceptibility measurements  \cite{Almeida:1997aa,Gama:1991aa,Henriques:1987aa}. They also agree with the experimental measurements of the anisotropy of the conductivity  estimated as of $\sigma_b/\sigma_a\approx 10^3$ \cite{Alcacer:1985aa}.
 
\begin{figure}[htbp]
\centering
\includegraphics[width=9cm]{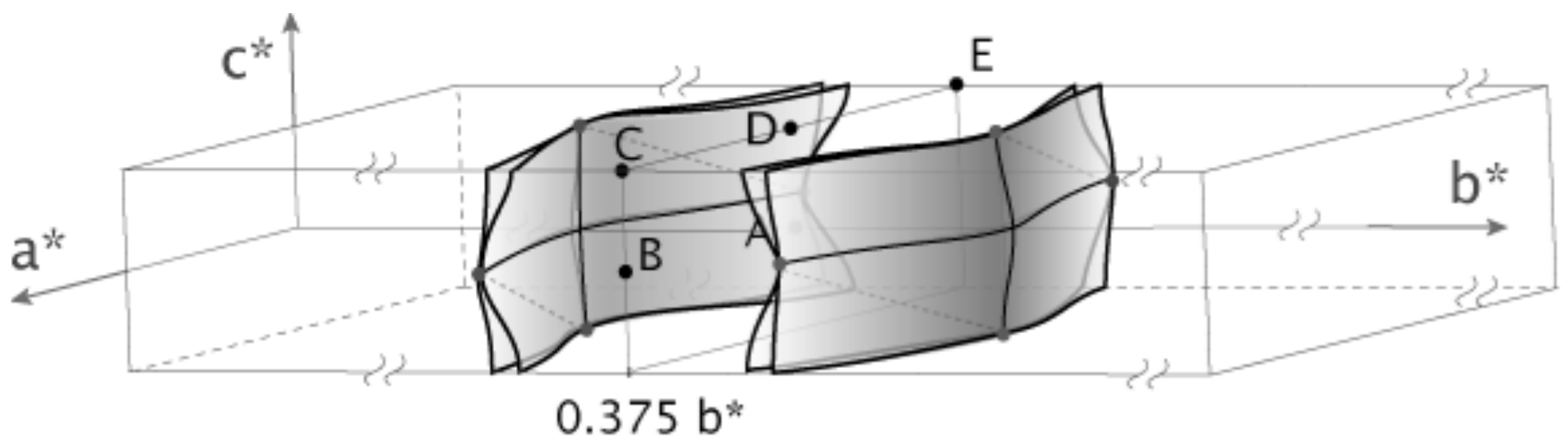}  
\caption{Estimated Fermi surface of the perylene chain in {(Per)$_2$[Pt(mnt)$_2$]}. Note that this is a largely enlarged picture around {$k_F= 0.375\ b^*$}.  (After reference \cite{Canadell:2004aa}).}
\label{Fig_6}
\end{figure}

 Since there are four perylene stacks per unit cell, there must be four HOMO bands crossing the Fermi level. If we would neglect the transverse interactions, the Fermi surface would be the superposition of four planes at $\pm k_F= \pm 0.375\ b^*$,  (by convention, $(b\cdot b^*) = 2 \pi$, $b=4.1891$~\AA; $k_F=\frac{1}{4} b^*=\frac{\pi}{2b}= 0.375$~\AA$^{-1}$ along $b^*$), but because of these transverse interactions, the Fermi surface splits into four sheets with some warping as shown in Figure \ref{Fig_6}. The warping of the  sheets  is very small,  of the order of 2 meV  along $a^*$ and is even smaller along $c^*$. 

With these  weak interchain interactions, the four Fermi surface sheets may touch each other, and by hybridization between
them there could be regions with closed pockets. Although these effects usually would be ignored, they may become important at very low temperatures.
The geometry of the Fermi surface was also tested with detailed measurements of angular effects in the magnetoresistance, but the results are difficult to interpret because there are simultaneously angular effects and orbital effects with quantum interference \cite{Graf:2009aa} (see below).
%
\subsection{Transport and Magnetic  Properties}
\begin{figure}[htbp]
\centering
\includegraphics[scale=0.9]{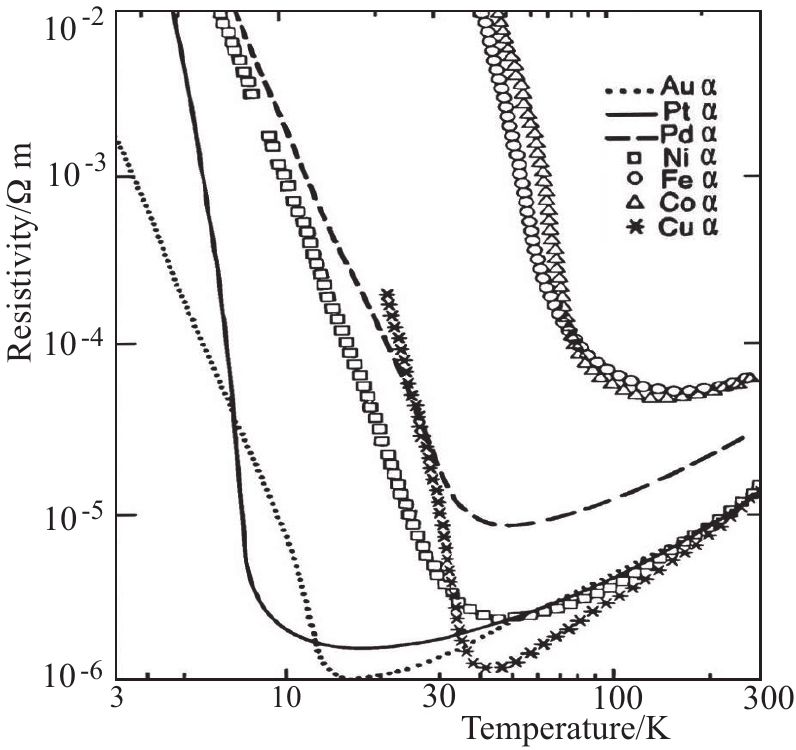}  
\caption{Resistivities as a function of temperature for the {$\alpha$}-phases  of the {(Per)$_2$[M(mnt)$_2$]}  famility (after references \cite{Henriques:1984aa,Alcacer:1985aa,Gama:1989aa,Gama:1991aa}).}
\label{Fig_7}
\end{figure}

The $\alpha$-(Per)$_2$[M(mnt)$_2$] (M=Pt, Au, Ni, Cu) compounds show room temperature conductivities of the order of 700 S/cm along the high conductivity $b$ axis and the anisotropy in the $a, b$ plane is estimated to be of the order of 900 - 1000 (see Figure \ref{Fig_7}). 

The room temperature conductivity of the Pd compound is of the order of 300  S/cm and the Co and Fe compounds (with a crystal structure where the $b$ axis is doubled) present a relatively lower conductivity. In all members of the $\alpha$ series, the resistivity, $\rho(T)$, decreases with the temperature, with the conductivities following a metallic regime as $\sigma\sim T^{-\alpha}$, with $\alpha=1.5$, reaches a minimum at a temperature $T_\rho$ and goes through a metal-insulator transition at $T_c$, corresponding to the maximum of $d\ln\rho/d(1/T)$. These results are compiled in Table \ref{tabela2}.
\begin{table}[htp]
\caption{Metal-insulator transition temperatures of the {$\alpha$-(Per)$_2$[M(mnt)$_2$]}.}
\begin{center}
\begin{tabular}{c|ccccccc}
\hline 
 M & Au & Pt &  Pd & Ni & Cu & Fe & Co\\
\hline 
$T_c/$~K  & 12   & 8   &  28 & 25 & 33 & 58 & 73  \\  
\hline   
\end{tabular}
\end{center}
\label{tabela2}
\end{table}
 The thermopower, $S$, in the metallic regime, is very similar for all the $\alpha$-phases and in the range of  32 - 42~$\mu$V~K$^{-1}$ at room temperature, and it decreases  upon cooling, as expected for metals, and in agreement with the resistivity data. At the metal-insulator transition temperature, $T_c$, $dS/d(1/T)$ also shows an anomaly and, at lower temperature,  the thermopower varies as $1/T$, as expected for a semiconductor. The positive sign of the thermopower in the metallic state is indicative of \emph{hole} type conduction, which is consistent with a $3/4$ filled band of the perylene chains. From the linear regime at high temperatures, it is possible to estimate a bandwidth, $W=4t$, of the order of $0.6$~eV for the series, with the exception of the Co and Fe compounds where it is of the order of $0.5$~eV. 

\begin{figure}[htbp]
\centering
\includegraphics[scale=0.9]{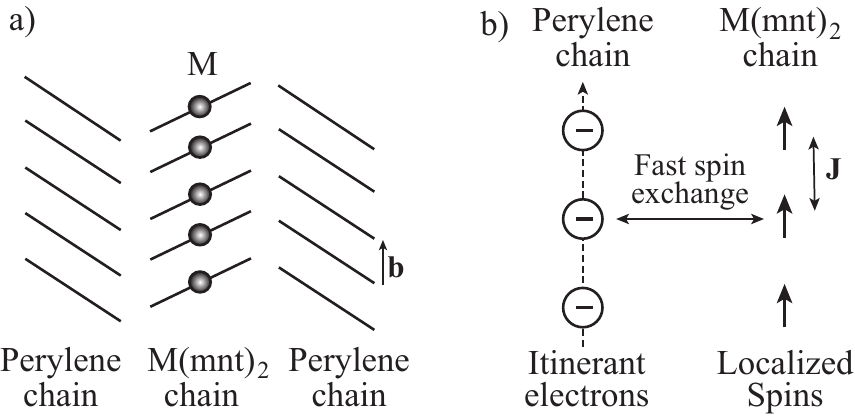}  
\caption{a) In {(Per)$_2$[M(mnt)$_2$]} (M=Ni, Pt, Pd), the crystal structures are formed by chains of {[M(mnt)$_2]$} units surrounded by perylene chains. b) Interactions along and between the chains: the perylene chain is a 3/4 filled band metal, while the {[M(mnt)$_2]$}  chain is an insulating array of localized spins, with fast spin exchange with the itinerant electrons. }
\label{Fig_8}
\end{figure}

The magnetic properties, namely the magnetic susceptibility, $\chi$,  and electron spin resonance, ESR, spectra have been studied in great detail in all the members of the series \cite{Alcacer:1974aa,Alcacer:1976aa,Alcacer:1980aa,Alcacer:1985aa,Henriques:1985aa}. It is important to mention that, through the combination of the susceptibility data with the ESR spectra, is was possible to separate the contributions to the susceptibility from both the itinerant electrons in the perylene chain and the localised electrons in the [M(mnt)$_2$] chain, for the paramagnetic species where M=Ni, Pd, Pt. The [Au(mnt)$_2$], being diamagnetic, was used as a term of comparison.  Figure \ref{Fig_8} schematically represents the general features of the magnetic properties for the salts with paramagnetic anions ($S=1/2$).\\ 
As we will discuss below in some detail, it is the particular combination of the structural and the electronic and magnetic properties and, in particular, the features of the Fermi surface, that make this series of quasi one-dimensional organic conductors so full of surprises. 
 %
\subsection{Charge Density Wave Formation. Precursor Effects}
\begin{figure}[htbp]
\centering
\includegraphics[scale=1]{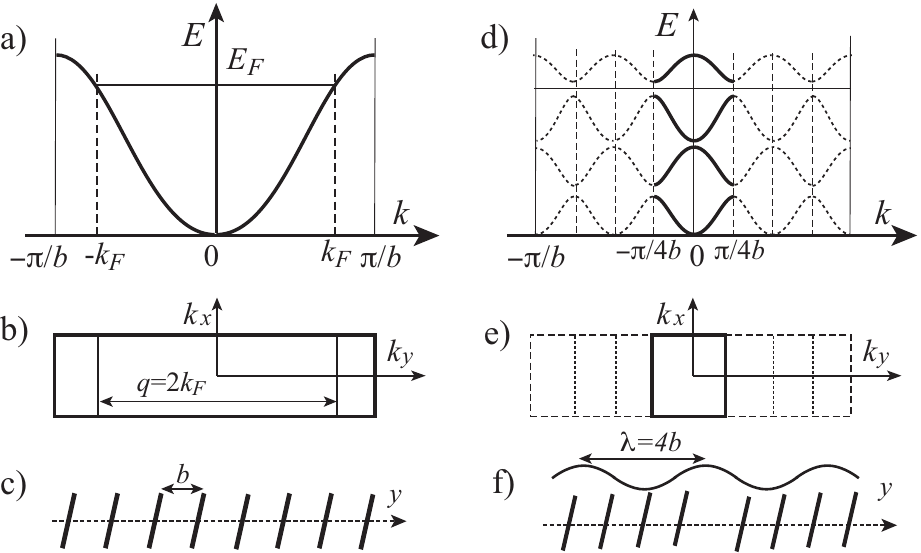}    
\caption{CDW formation for a {$3/4$}   filled band (1/4 filled hole band)  in a regular   {(Per)$^{0.5+}$} chain. a,b,c) Electron dispersion of the regular chain, Brillouin zone showing the {$q=2k_F$} phonon wave vector, and the lattice with cell parameter {$b$}, respectively. d,e,f) Opening of the gap at {$k_F=\frac{\pi}{4b}$}, CDW formation with {$\lam=4b$}, and tetramerization of the perylene chain.}
\label{Fig_9}
\end{figure}

As discussed above and illustrated in Figure \ref{Fig_2}, it would be expected to observe charge density wave formation and Peierls instabilities in the $\alpha$-phases of the (Per)$_2$[M(mnt)$_2$] family, as a result of the strong one-dimensionality evidenced by the crystal structure and high anisotropy in the conductivity. 
A schematic representation of the possible charge density wave formation and lattice modulation is shown in Figure \ref{Fig_9} 
for the $3/4$ (Per)$^{0.5+}$ filled band, which is expected to be unstable to tetramerization of the perylene chains (CDW). 

In fact, metal-insulator transitions are clearly observed in the plots of Figure \ref{Fig_7} for all members of the family, and the transition temperatures are given in Table \ref{tabela2}, where $T_c$ corresponds to the CDW formation (M=Au: $T_{CDW}=12$~K; M=Pt: $T_{CDW}=8$~K);  $2\Delta_0=\zeta k_B T$, with $5<\zeta<10$. In the Pt compound, (and for the Ni and Pd), as the tetramerization of the perylene chain occurs, there is a  simultaneous dimerization of the [Pt(mnt)$_2$] chains, corresponding to a spin-Peierls instability.   

\bigskip

\begin{figure}[htbp]
\centering
\includegraphics[width=5.5cm]{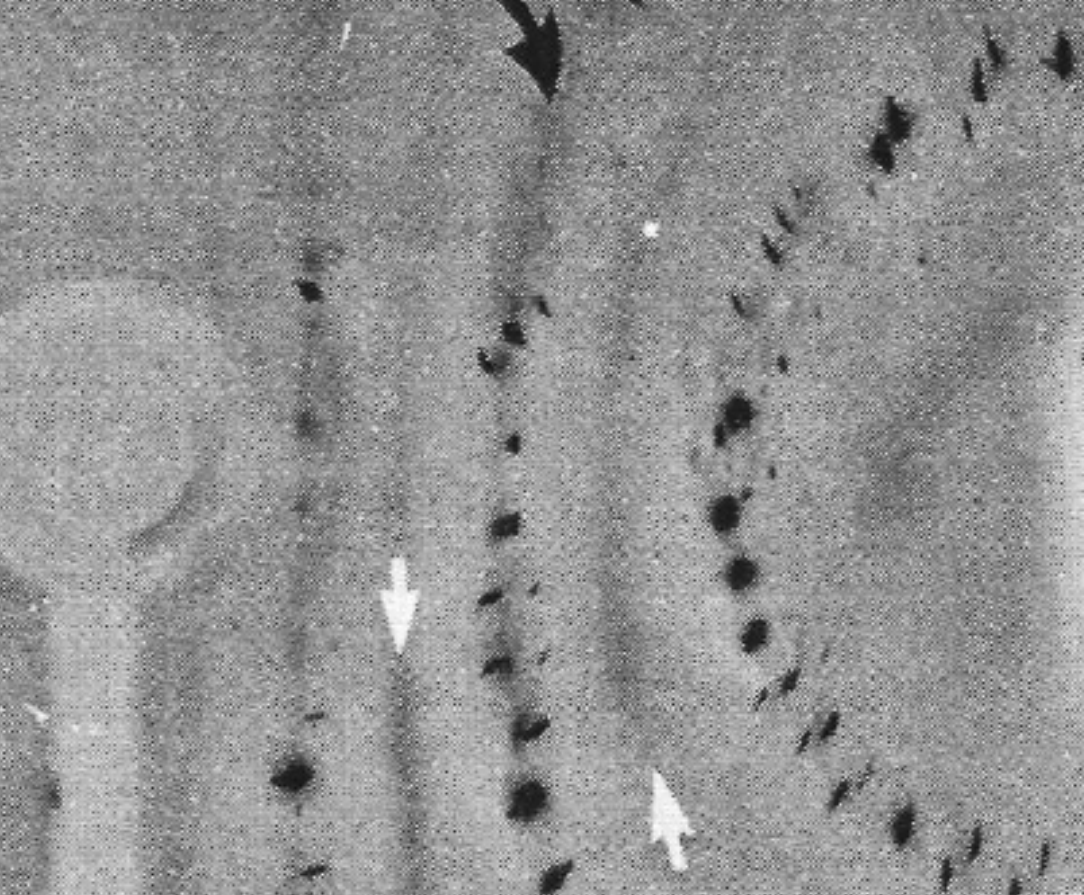}   
\caption{X-ray pattern of  {(Per)$_2$[Pt(mnt)$_2$]} showing, at 10 K, the broad {$\ha b^*$} diffuse scattering (white arrows) and the sharp {$0 b^*$} diffuse scattering (black arrow);  these lines appear below 25 K. On this pattern, the chain axis, {$b$}, is horizontal. This pattern is free from {$\lam/2$} contamination. After reference \cite{Henriques:1984aa}.}
\label{Fig_10}
\end{figure}

In order to study the expected structural instabilities of these 1D conductors, (see section 2.1 and Figure \ref{Fig_3}), Henriques \textit{et al}  \cite{Henriques:1984aa} performed X-ray diffuse scattering measurements in (Per)$_2$[M(mnt)$_2$] (M=Pt, Pd, Au), at several temperatures until reaching below the  observed metal-insulator transitions, with the exception of the Pt compound. These structural fluctuations are observed in the metallic state of (Per)$_2$[Pt(mnt)$_2$], as diffuse lines, at the  wave vector $\ha b^*$, as shown, at 10~K, in Figure \ref{Fig_10}, but start to appear at 25~K, well above the  critical temperature  $T_c=T_{CDW}=8$~K, which unfortunately could not be reached with the available experimental set-up. These diffuse scattering lines, attributed to the [M(mnt)$_2$] stacks, are also observed in  (Per)$_2$[Pd(mnt)$_2$], and, in this case, they condense in superstructure reflections below 28~K, characteristic of a spin-Peierls transition.\\   
A puzzling feature shown in (Per)$_2$[Pt(mnt)$_2$] (Figure \ref{Fig_10}), and not shown in (Per)$_2$[Pd(mnt)$_2$], is the presence of sharp diffuse lines passing through the main Bragg reflections, belonging to diffuse sheets with the reduced wave vector $q_b = 0 b^*$. Their origin could not be understood at the time. Another unexpected result of this investigation is that no diffuse lines were detected in (Per)$_2$[Au(mnt)$_2$] down to 12 K, contrary to  (Per)$_2$[Pt(mnt)$_2$] and (Per)$_2$[Pd(mnt)$_2$] which show 1D structural fluctuations below 25~K and 100~K respectively. 

The instabilities in  one-dimensional electronic systems may occur at  the {$q_b=2 k_F$} or the $q_b=4k_F$ wave vectors,
where $k_F$ is the Fermi wave vector. Through coupling with the phonons, these (CDW or SDW) instabilities may induce structural instabilities at the same wave vector.

Considering  that the conduction band (from the perylene chain) is 3/4 filled, corresponding to the formula (Per)$_2^+$[M(mnt)$_2$]$^-$  with one electron per dithiolate, we will be in the presence of a $q_b=2 k_F$ instability. This corresponds to $S=1/2$ for each [Pt(mnt)$_2$]$^-$ and [Pd(mnt)$_2$]$^-$ anion, and a closed-shell ion, with spin pairing ($S=0$) for the [Au(mnt)$_2$]$^-$ anion. This is in agreement  with the magnetic properties of these charge transfer salts, showing a Curie-Weiss law (at high temperature) for the Pt  \cite{Alcacer:1979aa,Alcacer:1980aa} and Pd  \cite{Alcacer:1976aa}  species, and no such behaviour for the Au species,  which show only a small temperature independent susceptibility typical of metals due to the perylene stacks \cite{Flandrois:1980aa}. 
  
 Similar diffuse scattering studies were done in (Per)$_2$[M(mnt)$_2$] (M=Cu, Ni, Co and Fe) \cite{Gama:1993aa}.  The metal-insulator transitions observed at 33, 25, 73 and 58~K for the Cu, Ni, Co and Fe salts, respectively, were found to correspond to the tetramerization on the perylene chains in the case of the Ni and Cu derivatives, and to the dimerization for the Fe and Co derivatives (where the $b$ axis was doubled). In all cases, the periodic lattice distortion can be associated to a $2k_F$ Peierls instability of the perylene chains. The Ni compound shows, in addition, a $4k_F$ distortion, corresponding to  the dimerization of the [Ni(mnt)$_2$]$^-$ chain of localized 1/2 spin, leading to a non magnetic spin-Peierls like ground state.

\medskip

A remarkable phenomenon, which was observed, was the  non-linear electrical  transport, both in broad brand and narrow band noise in the CDW state of (Per)$_2$[M(mnt)$_2$], (M=Au, Pt, Co), that  provided, for the first time in an organic conductor, clear evidence of coherent CDW motion as the origin of non-linear transport. These non-linear effects are denoted by a large increase of current above a critical threshold field, associated with an increase of broad and narrow band noise. In the Pt compound, the presence of narrow band noise with frequencies  proportional to the excess current was in excellent agreement with the coherent motion of a $2k_F$  CDW ($\lam=4b=16.8$~\AA) \cite{Lopes:1994aa,Lopes:1995aa,Lopes:1995ab,Dumas:1995aa,Lopes:1997aa}.
%
 \subsection{Nature of the Phase Transition in the Platinum Salt} 
\begin{figure}[htbp]
\centering
\includegraphics[scale=0.9]{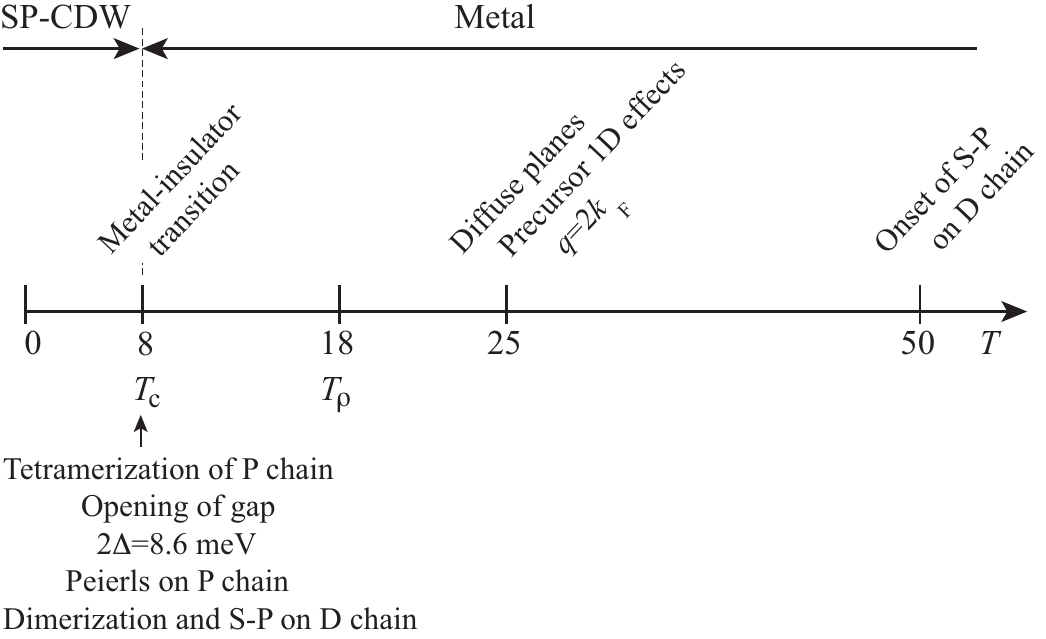}    
\caption{Cascade of phenomena (precursor effects and transition) for the {(Per)$_2$[Pt(mnt)$_2$]} compound as the temperature is lowered. The P chain is the perylene chain, while the D  chain  is the {[Pt(mnt)$_2$]} chain.}
\label{Fig_11}
\end{figure}

The most remarkable and unique feature of the  (Per)$_2$[M(mnt)$_2$] 1D solids is the fact that the perylene chain is a  quarter (hole) filled band metal, at room temperature, while the band in the [M(mnt)$_2$] chain is half filled, with localized spins for M=Ni, Pt and Pd (S=1/2), exhibiting fast spin exchange with the itinerant electrons on the perylene chain (see Figure \ref{Fig_8}). For M=Au, the chain is diamagnetic, making a good reference to study the peculiar magnetic properties of the series \cite{Alcacer:1980aa}. In this section, we will focus on the nature of the metal-insulator transition in the platinum compound, which has been the most studied (along more than 40 years) and is still  a matter of controversy. Figure \ref{Fig_11} illustrates the cascade of phenomena  observed in (Per)$_2$[Pt(mnt)$_2$], when going from room temperature to low temperatures.  

The perylene chain is metallic at room temperature and, at {$T\sim 8$~K}, undergoes a Peierls (CDW) transition to an insulating state, where it tetramerizes with wave vector $q_{Per} = \pi/2b$ and the opening of a gap of 8.6 meV. 
On the other hand, the [Pt(mnt)$_2$], (S = 1/2), chain is a Mott insulator that undergoes a spin-Peierls transition, dimerizing with wave vector $q_{Pt} = \pi/b$ and forming a spin singlet. An interesting aspect is that even though $q_{Pt} = 2q_{Per}$, diffuse X-ray scattering, specific heat, and electrical transport indicate that both the CDW and SP transitions occur at the same, or at very close temperatures, ($T_c=T_{SP\text{-}CDW}$) \cite{Henriques:1984aa,Gama:1993aa,Bonfait:1993aa}.
This is an indication that the  two chains are strongly coupled, even with the mismatch in the $q$ vectors. 

The transition is associated to precursor effects when coming from high to low temperatures: at {$T\sim 50$~K}, the magnetic susceptibility, $\chi$, is affected by the onset of the spin-Peierls instability on the dithiolate chain, while at $T\approx 25$~K, diffuse planes appearing on the X-ray diffraction pattern are a symptom of the displacive distortion at the phonon wave vector $q=2k_F$, $k_F$ being the Fermi wave vector for the electron dispersion (on the perylene chain). 
 
The nature of the transition,  and its precursor effects, has been studied again and again, using many experimental techniques, including electron spin resonance, magnetic susceptibility, transport properties (resistivity and thermopower) and X-ray diffuse scattering, as previously discribed, as well as specific heat \cite{Bonfait:1993aa} and nuclear magnetic resonance (NMR), both {$^1$H} \cite{Bourbonnais:1991aa,Green:2012aa} and  {$^{195}$Pt} NMR \cite{Green:2012aa}. 

\begin{figure}[htbp] 
\centering
\includegraphics[scale=0.9]{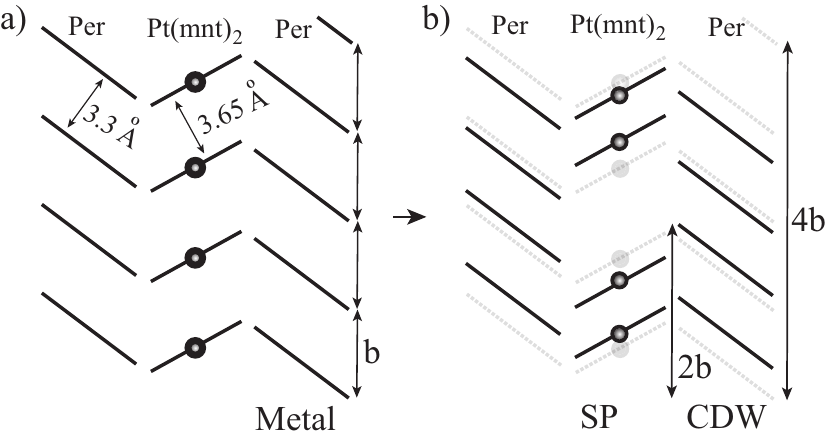} %
\caption{a) Configurations of the perylene chains (outer planes) and {[Pt(mnt)$_2$]} chain (inner plane) along {$b$}, in the metallic state ({$T>T_c$}). b) Configurations at {$T<T_c$}: Spin-Peierls (SP) in the {[Pt(mnt)$_2$]} chain (dimerized) and CDW in the perylene chain (tetramerized).}
\label{Fig_12}
\end{figure}

To make the story short, the present understanding based on, particularly, the comparison of the NMR data with all other results, one concludes that  the tetramerization on the perylene chain, corresponding to the  Peierls transition, 'forces' the dimerization on the [Pt(mnt)$_2$] chain, corresponding to a spin-Peierls transition (see Figure \ref{Fig_12}).   

\begin{figure}[htbp]
\centering
\includegraphics[scale=0.9]{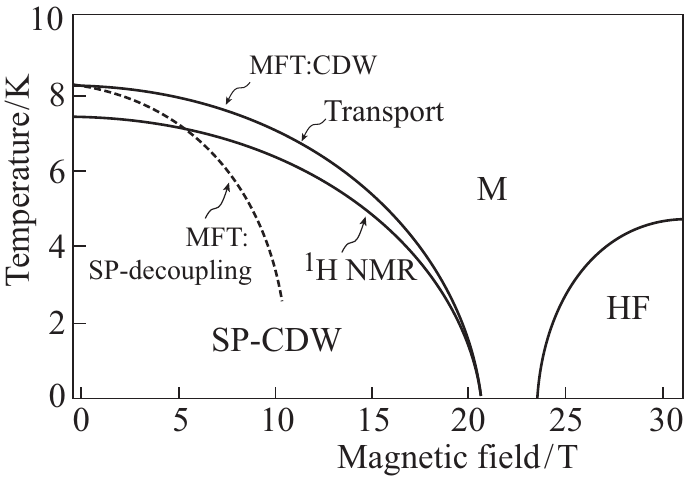}    
\caption{Temperature-magnetic field phase diagram of {(Per)$_2$[Pt(mnt)$_2$]} showing the boundaries for the spin-Peierls (SP), the charge density wave (CDW),  the paramagnetic-metallic (M), and the high field (HF) phases, derived from the onset of splitting of NMR spectra, and metal-insulator transitions from  electrical transport data, for a field applied along the {$a$}-axis. 'MFT:CDW' and 'MFT:SP-decoupling' are the mean-field theoretical predictions. (After references \cite{Green:2011aa,Green:2012aa}).}
\label{Fig_13}
\end{figure}

The details of the coupling between the Peierls transition in the perylene chain and the spin-Peierls transition in the [Pt(mnt)$_2$] is not yet clear, but it appears that a unique combination of SP and CDW order parameters arises.
Based on NMR results, in particular, on the  linear scaling of  $1/T_1$ \textit{vs.} \ $T\ \chi(T)$ \cite{Bourbonnais:1991aa}, which show that the spin degrees of freedom remain diffusive and weak until $T_{SP\text{-}CDW}$, it is not likely that the SP dimerization of [Pt(mnt)$_2$]  would be a result of quantum spin dynamics alone. However, electron paramagnetic resonance (EPR) studies indicate that an exchange interaction between the itinerant and the localized spins, associated with the onset of the $T_{SP\text{-}CDW}$ transition, might be involved in the cooperative SP-CDW transition. One \cite{Xavier:2003aa}  and two-chain \cite{Bourbonnais:1991aa} 1D Kondo lattice models have been considered, but at present there is no complete theory that treats the two-chain (Per)$_2$[Pt(mnt)$_2$] case. It has been shown \cite{Green:2011aa} that, even at high magnetic fields, this coupling persists, and that the initial tetramerization in the perylene chains may promote the stabilization of the SP dimerization and the resulting spin-singlet ground  state, in spite of the predictions of  mean-field theory \cite{Bray:1983aa}  that place the spin-Peierls decoupling well below the CDW phase boundary, as shown in Figure \ref{Fig_13}, which represents the temperature \textit{vs.} magnetic field phase diagram. 

Three regions of the (Per)$_2$[Pt(mnt)$_2$] phase diagram are shown: the metallic phase (M), above 8~K; the  SP-CDW phase, below 8~K and below 20 - 25~T, and the high-field (HF) phase, below 8~K and above 20 - 25~T. The SP and CDW order parameters remain coupled at high fields.  NMR spectra and spin relaxation data indicate that the full spin-Peierls state  forms below the CDW transition seen in transport measurements, at least, at low magnetic fields.  The CDW ground state formation appears to be a necessary condition to drive the formation of the spin-Peierls ground state, and the commensurability (2:1) of the two order parameter periodicities is likely to favour this effect.
%
\subsection{Perylene-Metal Dithiolates Under Extreme Conditions}   
The behaviour of matter under \emph{extreme conditions}, meaning, in particular, very low temperatures, high pressures and  high magnetic fields is of the greatest importance, for both theoretical and technological development. The history of  superconductivity is a good example. 

\medskip

The search for superconductivity in organic solids led to the study of the transport properties under high pressure, based on the belief that forcing the molecules in a chain to come closer would lead to higher transfer integrals and eventually to a superconducting phase. This was actually achieved with the observation of superconductivity,  for the first time in an organic solid, namely (TMTSF)$_2$PF$_6$, \cite{Jerome:1980aa}, the first of the series of the so called Bechgaard superconducting salts.  

Pressure effects on the transport properties of (Per)$_2$[M(mnt)$_2$] for M=Pt and Au were first studied by Henriques \textit{et al}, \cite{Henriques:1986aa}, at pressures up to 27 kbar. It was observed that the room temperature conductivity measured along the needle axis, ($b$-axis), increases with pressure and tends to saturate above approximately 9 kbar. The results show that the evolution of the electronic correlations taking place in the neighbouring conducting chains was quite unusual, when compared with  other 1D organic conductors, and further studies were needed.

\begin{figure}[htbp] 
\centering
\includegraphics[scale=0.9]{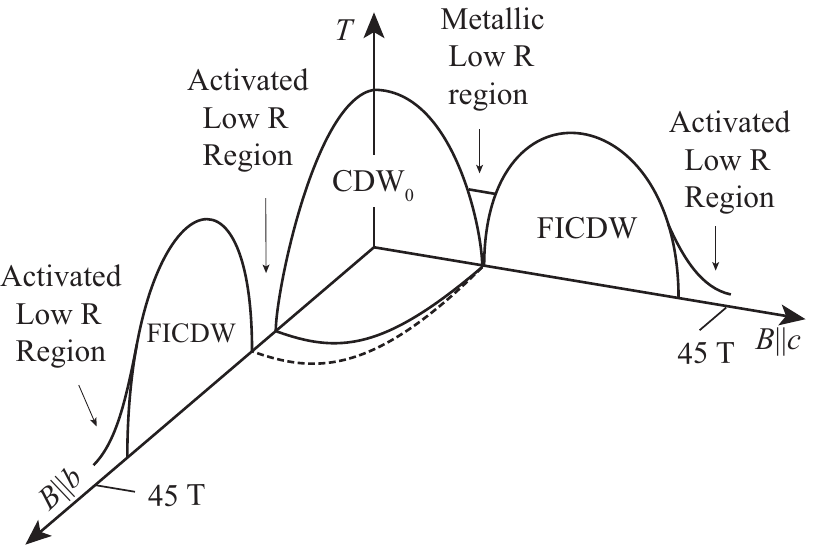}  
\caption{Schematic phase diagram of {(Per)$_2$[Pt(mnt)$_2$]}, where field induced charge density wave (FICDW) phases are shown at magnetic fields up to 45 T.  (After reference \cite{Brooks:2006aa})}
\label{Fig_14}
\end{figure}

The low  $T_c$ values of the Peierls transitions in these compounds (8~K in Pt and 12~K in Au) elect them as ideal systems to test the effect of high magnetic fields on a CDW. Initial measurements up to 8~T  \cite{Bonfait:1991aa}) detected, for the first time in a CDW system, a sensible decrease of the transition temperature with magnetic field. Larger effects were latter detected in measurements up to 18~T which, although not large enough for a complete suppression of the CDW, put into evidence a larger and anisotropic dependence in the Pt compound, which was ascribed to the coupling to the magnetic chains undergoing a spin-Peierls transition \cite{Matos:1996aa}. The total suppression of the  CDW was finally observed in 2004 \cite{Graf:2004ab} when studying the high field behaviour of the CDW ground state in several of the elements of the series. At very high magnetic fields (up to 45~T) a cascade of transitions in the platinum compound, tentatively ascribed to field induced charge density wave (FICDW) phases, was discovered \cite{Brooks:2006aa}. Figure \ref{Fig_14} schematically represents the phase diagram of the platinum compound, where field induced charge density wave (FICDW) state phases are shown at high magnetic fields (up to 45 T).

The theory of CDW under high magnetic fields is still incomplete, but considerable   achievements have been reached, namely for (Per)$_2$[Pt(mnt)$_2$], particularly by Lebed \textit{et al} \cite{Lebed:2003aa,Lebed:2007aa,Lebed:2007ac,Lebed:2009aa}.
\subsection{Quantum Interference Effects}
The interference of electrons was first observed in 1927 by  Davisson and Germer \cite{Davisson:1927aa}  in the diffraction pattern of electrons incident on nickel, and was the confirmation of the de Broglie hypothesis that electrons  behaved as waves, with wave length $\lam=h/p$ ($p=mv$). A double slit  experiment with electrons was first reported by J\"{o}nsson in 1961 \cite{Jonsson:1961aa}. Quantum interference of electron waves in a metal has been observed by Stark and Friedberg \cite{Stark:1971aa} and later by Sandesara and Stark \cite{Sandesara:1984aa}, and is known as the Stark interference effect. This effect is observed as large amplitude oscillations in the transverse magnetoresistance signal, periodic in inverse field, which result from the direct interference of normal-state electron waves, corresponding to  two electron ''trajectories'' on different Fermi-surface sheets, with a phase difference of $\Psi=2\pi\phi/\phi_0$, where $\phi/\phi_0$ is the ratio of the magnetic flux enclosed between the electron paths to the magnetic flux quantum $\phi_0=h/e$. 

\begin{figure}[htbp]
\centering
\includegraphics[scale=0.9]{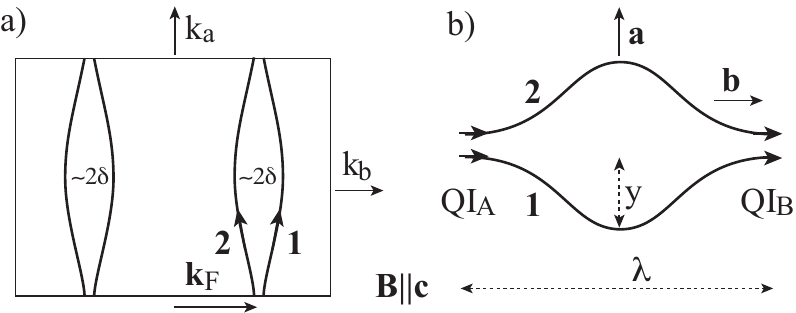}    
\caption{a) Simplified Fermi surface for  {(Per)$_2$[Au(mnt)$_2$]} \cite{Canadell:2004aa}. The shortest lattice parameter, $b$, is of the order of 4.19~\AA \ and the corresponding bandwidth, $t_b$, is 150~meV, the other being much smaller. b) Real-space motion of the carriers with wavelength approx. 2.5/B~micron tesla and amplitude approx.  50/B~nm~T. Transmission or reflection of the carriers can occur at the interference nodes {$\mathrm{QI_A}$} and {$\mathrm{QI_B}$}. (After reference \cite{Graf:2007aa}).} 
\label{Fig_15}
\end{figure}

The Stark interference effect has been observed through magnetotransport measurements  in the  one dimensional conductor {(Per)$_2$[Au(mnt)$_2$]} by suppressing the charge density wave state with pressure (above 3~kbar), below the transition temperature $T_{CDW}=12$~K \cite{Graf:2007aa}, see Figure \ref{Fig_15}.\\
The quantum interference (QI) oscillation amplitude exhibits a temperature dependent scattering rate, indicative of an inhomogeneous metal CDW ground state. The QI oscillation frequency reveals a Fermi-surface topology of the unnested Fermi surface, in agreement with that recently computed by Canadell \textit{et al} \cite{Canadell:2004aa} and schematically shown in Figure \ref{Fig_6}. 
\subsection{Superconductivity at Last}
\begin{figure}[htbp]
\centering
\includegraphics[scale=0.9]{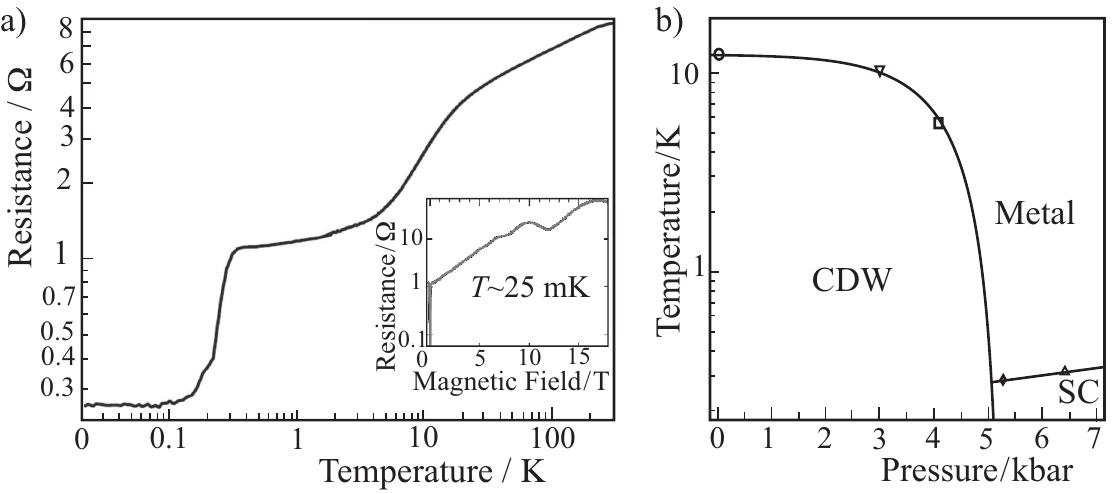} %
\caption{a) Resistance vs. temperature for {(Per)$_2$[Au(mnt)$_2$]} at {$5.3$~kbar}. Inset: magnetoresistance, showing the SC critical field and Stark effect. b) CDW-SC phase diagram (after reference \cite{Graf:2009ab}).}
\label{Fig_16}
\end{figure}

Finally, there is the discovery of superconductivity in the Au compound, with the peculiarity that it occurs in the vicinity of the CDW state \cite{Graf:2009ab}. As shown in Figure \ref{Fig_16}a), the resistance of (Per)$_2$[Au(mnt)$_2$], under a pressure of  5.3kbar, drops sharply below 0.35 K, indicative of the onset of superconductivity, although a residual resistance remains in the sample. In the inset of Figure \ref{Fig_16}~a), the magnetoresistance at 25~mK is displayed showing, at low fields, the critical-field behaviour characteristic of superconductivity, and at higher fields the oscillations characteristic of the Stark interference. 

(Per)$_2$[Au(mnt)$_2$] has a non-magnetic anion, and exhibits a Peierls type transition to a CDW ground state below 12~K \cite{Almeida:1997aa}. The resulting insulating CDW state is suppressed in high magnetic fields \cite{Graf:2004aa,Brooks:2006aa}. The CDW state can also be suppressed under high pressure, but, under some conditions, a residual ''quantum melted'' CDW character remains \cite{Mitsu:2005aa}. Under pressures above 3~kbar, the magnetoresistance shows a metallic state with long range order, exhibiting a quantum interference Stark effect. 
The onset of superconductivity occurs at $\approx 300$~mK, and the upper critical field (perpendicular to the perylene stacking axis) is 50~mT. Although to fully demonstrate superconductivity will require measurements, such as heat capacity,  or magnetic susceptibility, which are difficult to  do under high pressure, the critical field phase diagram suggests BCS-like superconducting behaviour, (see figure 3 of reference \cite{Graf:2009ab}) with the parameters $B_0=49.4$~mT and $T_c =0.313$~K. 
Figure \ref{Fig_16}~b) schematically shows the proposed CDW-SC phase diagram. The metal-CDW transitions were determined from the standard $d\ln(R)/d(1/T)$ peaks on the resistance, and the metal-superconductor transition from the onset of superconductivity.  An interesting aspect in this compound  is that superconductivity emerges from a charge density wave state, at variance with most other superconductors, where superconductivity  emerges from antiferromagnetic states.  
%
\section{Concluding Remarks}
The enthusiastic expectations of the early nineteen seventies did not bring about room temperature superconductors, but led to new theoretical developments in the science of condensed matter and to promising emergent technologies. On the scientific side, the results have shown how impressive is the physics dealing with the motion of electrons in one dimension. In the particular case of the two-chain highly one-dimensional conductors of the (Per)$_2$[M(mnt)$_2$] family, the work done, along these more than 40 years, has shown how rich in physical phenomena are these amazing materials. The story told  in this paper is far from exhaustive and it is certainly unfinished---there will be more surprises as theoretical  and experimental methods progress. There are still no theoretical answers to many questions, such as, for example, the nature of the mechanism of coupling of the charge density wave (CDW) and the spin-Peierls (SP) state in the perylene and in the [M(mnt)$_2$] chains, respectively, and the origin of the field induced spin density wave (FISDW) state under high magnetic fields. 

There were many actors in this story, which includes several PhD thesis. The development of the research in organic conductors, in Portugal, benefited from several close collaborations, mainly with the groups of Denis J\'{e}r\^{o}me and Jean-Paul Poujet  at the Universit\'{e} Paris-Sud at Orsay and the group of the late James Brooks, from the National High Magnetic Field Laboratory (NHMFL) in Tallahasse (USA). As for the best part of the theoretical work, it is due to the precious contribution of Claude Bourbonnais from the Universit\'{e} de Sherbrooke in Canada and also to Enric Canadell from Institut de Ci\`{e}ncia de Materials de Barcelona (ICMAB-CSIC). A word of recognition is also due to the many co-authors who contributed for the progress of research in this family of compounds.

\section*{Acknowledgements}
This work was supported by FCT-Portugal  contracts UID/EEA/50008/2013 and UID/Multi/04349/2013.

\bibliographystyle{unsrtnt}
\bibliography{Electrons1D}

\end{document}